\documentclass[12pt,a4paper]{article}

\usepackage[margin=2.5cm]{geometry}
\usepackage{authblk}
\usepackage[dvipsnames]{xcolor}
\usepackage[unicode]{hyperref}
\hypersetup{
  colorlinks=true,
  citecolor=MidnightBlue,
  linkcolor=MidnightBlue,
  urlcolor=MidnightBlue,
  linktocpage=true
}
\makeatletter
\newcommand{\anchorlabel}[2]{\phantomsection\def\@currentlabel{#2}\label{#1}}
\makeatother
\usepackage{tabularx}
\usepackage{graphicx}
\usepackage{amssymb,amsmath,bm,tensor,braket}
\usepackage[varg]{txfonts}
\usepackage{enumerate}
\usepackage{mathtools}
\usepackage{tensor}
\usepackage[capitalize]{cleveref}
\usepackage[utf8]{inputenc}
\usepackage[normalem]{ulem}
\usepackage{dcolumn}
\usepackage{soul}
\usepackage{subfigure}
\usepackage{booktabs}
\usepackage{comment}
\usepackage{siunitx}
\usepackage{mathrsfs}
\usepackage[T1]{fontenc}
\usepackage{amsmath}
\usepackage{float}
\usepackage{adjustbox}

\newcommand{\ii}{\mathrm{i}}
\newcommand{\ee}{\mathrm{e}}

\title{Pair-Dependent Drift of Kerr Neighboring-Overtone Gap Minima}

\author[1]{Yuye Wu}
\author[2,3,4]{Hong-Bo Jin\thanks{Corresponding author. E-mail: \texttt{hbjin@bao.ac.cn}}}
\affil[1]{Department of Astronomy, College of Physical Science and Technology, Xiamen University, Xiamen 361005, China}
\affil[2]{National Astronomical Observatories, Chinese Academy of Sciences, Beijing 100101, China}
\affil[3]{School of Astronomy and Space Science, University of Chinese Academy of Sciences, Beijing 100049, China}
\affil[4]{The International Center for Theoretical Physics Asia-Pacific (ICTP-AP), University of Chinese Academy of Sciences, Beijing 100190, China}

\date{}

\begin{document}

\maketitle

\begin{abstract}
We study adjacent Kerr quasinormal-mode overtones under a spin scan with overtone labels held fixed, using a public Leaver-type solver on a uniform grid. The observable is the modulus of the complex-frequency separation between neighbors; its minima are analyzed through the spin derivative of the squared separation, which supplies a smooth real diagnostic without differentiating the modulus itself. Clear interior minima appear, but their spin locations shift between neighboring pairs even within one \((s,\ell,m)\) sector and align with dominant zeros of the diagnostic and with radial turning of the separation vector in the complex-frequency plane. Representative extra sectors and smooth no-trigger cases support selectivity. Minimum drift is naturally read as drift of that dominant zero; the language connects to complex-spectral pole proximity for Kerr flows without identifying each minimum with an exceptional-point coalescence or claiming a universal rule over the full spectrum.
\end{abstract}


\noindent \textbf{\em Introduction.}
Quasinormal modes play a central role in black-hole perturbation theory, ringdown physics, and black-hole spectroscopy~\cite{ChandrasekharDetweiler1975,Chandrasekhar1985,BertiCardosoStarinets2009,BertiEtAl2016KerrSpectroscopy,BertiEtAl2025Spectroscopy}.
In the Kerr case, the QNM spectrum is also known to exhibit rich structure. Examples include branching and bifurcation near extremality, resonant-frequency patterns in rapidly rotating black holes, resonant excitation, nontrivial excitation factors, eigenvalue repulsion, and exceptional-point-related behavior~\cite{YangEtAl2013,YangEtAl2013b,Cardoso2004Rapid,Hod2011NearExtremal,Motohashi2025,YangBertiFranchini2025,LoSabaniCardoso2025,DiasEtAl2022,CavalcanteEtAl2024PRL,CavalcanteEtAl2024PRD,SantosEtAl2026}.
For broader exceptional-point physics beyond the immediate Kerr setting, see also Ref.~\cite{Heiss2012}.
Numerical-relativity analyses further show that overtones can matter already near the peak strain amplitude~\cite{LondonEtAl2014,BhagwatEtAl2018StartTime,GieslerEtAl2019,Bhagwat2020}.
Excitation studies suggest that, at intermediate and high spins, several higher overtones can be efficiently excited, while some exhibit anomalous behavior close to extremality~\cite{BertiCardoso2006,DorbandEtAl2006,ZhangBertiCardoso2013,Oshita2021,OshitaCardoso2024,LoSabaniCardoso2025}.
These results motivate local organizing principles in the complex-frequency plane rather than treating the spectrum as isolated scalar gaps; see also Ref.~\cite{JaramilloPanossoMacedoAlSheikh2021}.

Less direct, however, is how \emph{neighboring overtones} approach and separate under varying Kerr spin when each mode is continuously tracked at fixed overtone label~\cite{Motohashi2025,YangBertiFranchini2025,LoSabaniCardoso2025,YangEtAl2013,YangEtAl2013b}.
Accordingly, we study the raw complex-frequency separation between adjacent overtones along a one-parameter scan in \(a\).
A clear pattern emerges: different neighboring pairs develop interior minima at different spins, even within one \((s,\ell,m)\) sector.
The task is to identify what sets these minima and why their locations drift between pairs.

Our account is local and geometric: minimum location becomes a zero-setting problem for the complex separation \(Z(a)\). The sampled minimum is governed by
\begin{equation}
F(a)=\Re\!\big(\overline{Z(a)}\,Z'(a)\big)\approx 0,
\end{equation}
so that pair-dependent minimum drift becomes the drift of the dominant zero-crossing location of \(F(a)\).

In the complex plane, the minimum is a radial turning event of the separation vector---the radial projection passes through zero while angular motion may persist---so the physics is carried by local separation dynamics, not by the scalar gap curve alone.

We then test the picture on a deliberately restricted representative set; triggered cases share the same local mechanism, but crossing details remain case-dependent.
Section~\ref{sec:pheno-triggers} (within the Discussion, section~\ref{sec:discussion}) outlines ringdown-oriented consequences and an empirical trigger taxonomy from the appendix scans.

The article follows a standard IMRaD layout (Introduction, Methods, Results, and Discussion): \emph{Introduction} above; \emph{Methods} (section~\ref{sec:methods}) for the label-fixed gap, diagnostic, and numerical protocol; \emph{Results} (section~\ref{sec:results}) for mainline minima, local and full-scan diagnostics, and representative validation; and \emph{Discussion} (section~\ref{sec:discussion}) for spectral interpretation, phenomenology, and conclusions.

\noindent \textbf{\em Methods.}
\anchorlabel{sec:methods}{II}
The methods split into a mathematical specification of the label-fixed neighboring-pair gap and the spin derivative diagnostic \(F\) (subsection~\ref{subsec:methods-framework}), and a concrete numerical implementation for Kerr scans (subsection~\ref{subsec:methods-numerical}).

\medskip
\noindent\textbf{Mathematical framework (label-fixed gap and diagnostic).}
\anchorlabel{subsec:methods-framework}{M1}
\par\smallskip
Throughout, each overtone is tracked in Kerr spin \(a\) with its label held fixed. For a neighboring pair \((n,n+1)\), the complex-frequency separation in the plane is measured by the modulus of the frequency difference:

\begin{equation}\label{eq:gap_pair}
  g_{n,n+1}(a)=\left|\omega_n(a)-\omega_{n+1}(a)\right|.
\end{equation}
The sampled spin location of the minimum gap is denoted by

\begin{equation}\label{eq:a_star}
  a^*_{n,n+1}=\operatorname{argmin}_a g_{n,n+1}(a).
\end{equation}
This raw neighboring-pair gap is the primary numerical input in what follows.

The same data are packaged as a complex separation vector,

\begin{equation}\label{eq:Z_def}
  Z(a)\equiv \omega_{n_2}(a)-\omega_{n_1}(a),
\end{equation}
where \((n_1,n_2)\) denotes the neighboring pair. The gap is simply its magnitude:
\begin{equation}\label{eq:gap_Z}
  g(a)=|Z(a)|,
  \qquad
  g^2(a)=|Z(a)|^2.
\end{equation}

Because \(g(a)\ge 0\), minimizing \(g(a)\) is equivalent to minimizing \(g^2(a)\). Working with \(g^2\) is also cleaner. Differentiating \(g(a)=|Z(a)|\) introduces a denominator \(1/|Z(a)|\). Differentiating \(g^2(a)=|Z(a)|^2\) avoids it.

The central local diagnostic in this work is

\begin{equation}\label{eq:F_def}
  F(a)\equiv \Re\!\left(\overline{Z(a)}\,Z'(a)\right).
\end{equation}

This diagnostic is directly tied to how the gap changes with spin. From equation~(\ref{eq:gap_Z}),
\begin{equation}
  g^2(a)=|Z(a)|^2=Z(a)\,\overline{Z(a)}.
\end{equation}
Differentiating with respect to \(a\) gives
\begin{equation}
  \partial_a g^2(a)
  =
  Z'(a)\,\overline{Z(a)}
  +
  Z(a)\,\overline{Z'(a)}
  =
  2\,\Re\!\left(\overline{Z(a)}\,Z'(a)\right).
\end{equation}

\begin{equation}\label{eq:dg2_master}
  \partial_a g^2(a)=2\,\Re\!\left(\overline{Z(a)}\,Z'(a)\right)=2F(a).
\end{equation}

Accordingly, a local minimum is characterized by the near-vanishing condition

\begin{equation}\label{eq:min_condition}
  F(a^*)\approx 0.
\end{equation}

This is the local statement used throughout the analysis.

\medskip
\noindent\textbf{Numerical implementation (label-fixed Kerr scan).}
\anchorlabel{subsec:methods-numerical}{M2}
\par\smallskip
Mainline scans use the public \texttt{qnm} package~\cite{Stein2019}, which supplies Kerr QNM frequencies via a Leaver-type continued-fraction solver with angular spectral treatment and cached interpolation~\cite{Leaver1985,CookZalutskiy2014}.
The focus is \((s,\ell,m)=(-2,2,2)\) and two neighboring overtone pairs, \(\mathrm{pair}(4,5)\) and \(\mathrm{pair}(5,6)\), at fixed overtone labels along the scan.

A coarse raw scan is performed on a uniform spin grid
\[
a\in[0,\,0.99],
\qquad
\Delta a_{\rm coarse}=0.01.
\]
At each grid point, the complex separation is constructed directly from the two resolved mode frequencies,
\[
Z(a)=\omega_{n_2}(a)-\omega_{n_1}(a),
\qquad
g(a)=|Z(a)|.
\]
The corresponding coarse sampled minima occur at
\[
a^*_{4,5}=0.85,
\qquad
a^*_{5,6}=0.90,
\]
with
\[
g_{\min}^{4,5}=0.120505,
\qquad
g_{\min}^{5,6}=0.068794.
\]

If the solver fails or returns a non-finite value, we record the point as unresolved. In practice, the resolved coarse scan extends up to \(a=0.99\).
These scans establish robust interior minima and motivate the analysis in section~\ref{sec:results}.

\noindent \textbf{\em Results.}
\anchorlabel{sec:results}{III}
This section reports the mainline minimum locations (subsection~\ref{subsec:results-mainline}), the local Cartesian and complex-plane structure at those minima (subsection~\ref{subsec:results-local}), and full-scan consistency together with zero-crossing alignment and extended validation (subsection~\ref{subsec:results-global}).

\medskip
\noindent\textbf{Mainline interior minima and pair-dependent drift.}
\anchorlabel{subsec:results-mainline}{R1}
\par\smallskip
Consider the mainline sector
\begin{equation}
  (s,\ell,m)=(-2,2,2),
\end{equation}
with \(\mathrm{pair}(4,5)\) and \(\mathrm{pair}(5,6)\) at fixed labels. The aim here is to establish that spin drift of the gap minimum is a robust feature of the raw neighboring-pair curve, not a grid artifact.
The emphasis on higher neighboring overtones matches ringdown and excitation work where intermediate and high overtones are prominent~\cite{LondonEtAl2014,BhagwatEtAl2018StartTime,GieslerEtAl2019,Bhagwat2020,BertiCardoso2006,DorbandEtAl2006,ZhangBertiCardoso2013,Oshita2021,OshitaCardoso2024}.

\begin{table}[htbp]
  \centering
  \caption{Mainline minimum summary for the direct raw neighboring-pair gaps (with overtone labels held fixed along the scan) in the $(s,\ell,m)=(-2,2,2)$ case.}
  \label{tab:mainline_minima_summary}
  \footnotesize
  \setlength{\tabcolsep}{4pt}
  \begin{tabular}{lcccc}
    \toprule
    Pair & $a^*_{\rm coarse}$ & $a^*_{\rm refined}$ & $g_{\min}^{\rm coarse}$ & $g_{\min}^{\rm refined}$ \\
    \midrule
    $\mathrm{pair}(4,5)$ & 0.850 & 0.850 & 0.120505 & 0.120505 \\
    $\mathrm{pair}(5,6)$ & 0.900 & 0.897 & 0.068794 & 0.066674 \\
    \bottomrule
  \end{tabular}
\end{table}

Table~\ref{tab:mainline_minima_summary} lists coarse and refined minima: both pairs show clear interior minima, but at different spins, so even in \((s,\ell,m)=(-2,2,2)\) there is no common minimum location.

The separation in minimum spin between the two pairs is
\begin{equation}
  \Delta a^* \equiv a^*_{5,6}-a^*_{4,5}=0.050 \quad \text{(coarse)},
\end{equation}
and
\begin{equation}
  \Delta a^* = 0.047 \quad \text{(refined)}.
\end{equation}
Coarse and refined scans agree, so the offset is not a fragile grid artifact but a stable difference between pairs.
Both minima are interior (two-sided valley, stable under refinement).

\medskip
\noindent\textbf{Local Cartesian and complex-plane behaviour at the sampled minima.}
\anchorlabel{subsec:results-local}{R2}
\par
\begin{figure}[t]
  \centering
  \includegraphics[width=\linewidth]{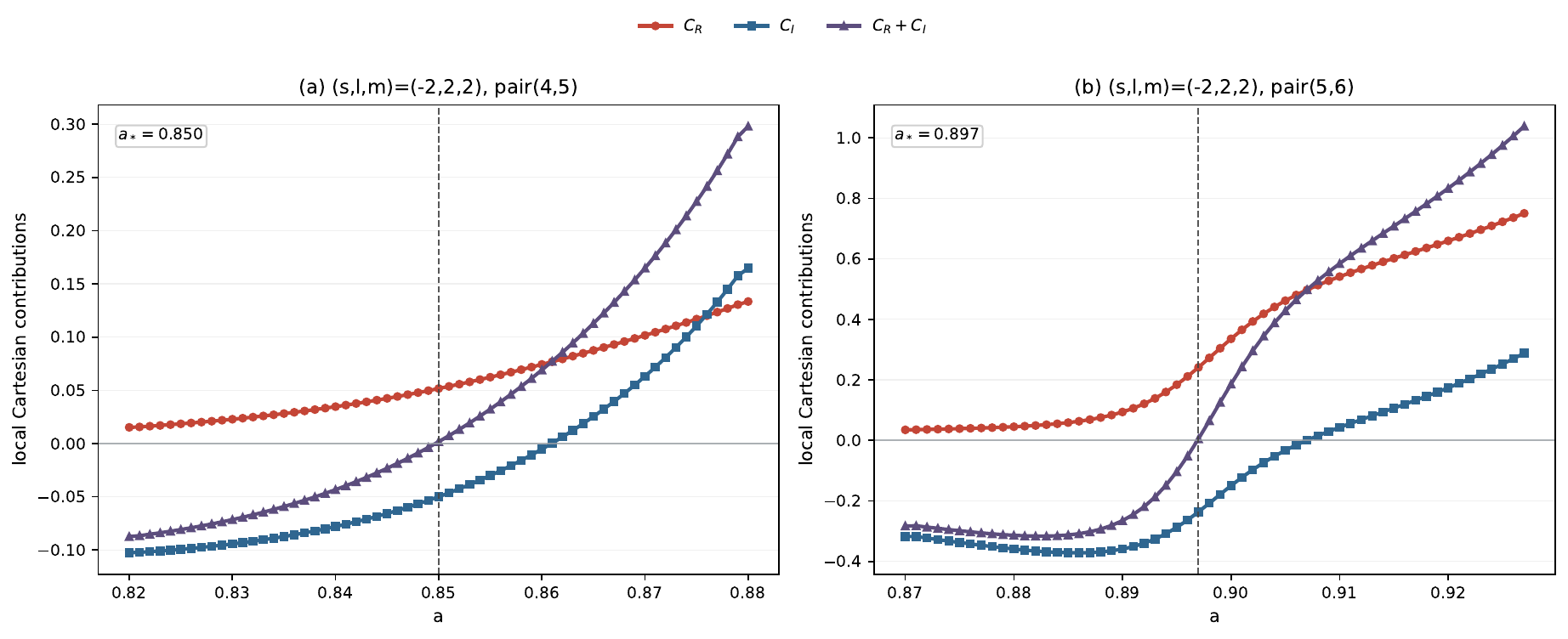}
  \caption{Local Cartesian balance near the sampled minima for the mainline Kerr case $(s,\ell,m)=(-2,2,2)$. In each panel, the real and imaginary contributions $C_R(a)$ and $C_I(a)$ are plotted together with their sum $C_R(a)+C_I(a)=\partial_a g^2(a)$. On the left side of the minimum the sum remains negative, near the minimum the two contributions nearly cancel, and on the right side the sum turns positive. The dashed vertical line marks the sampled minimum location.}
  \label{fig:mainline_local_cartesian_balance}
\end{figure}

\begin{figure}[!t]
  \centering
  \includegraphics[width=0.95\linewidth]{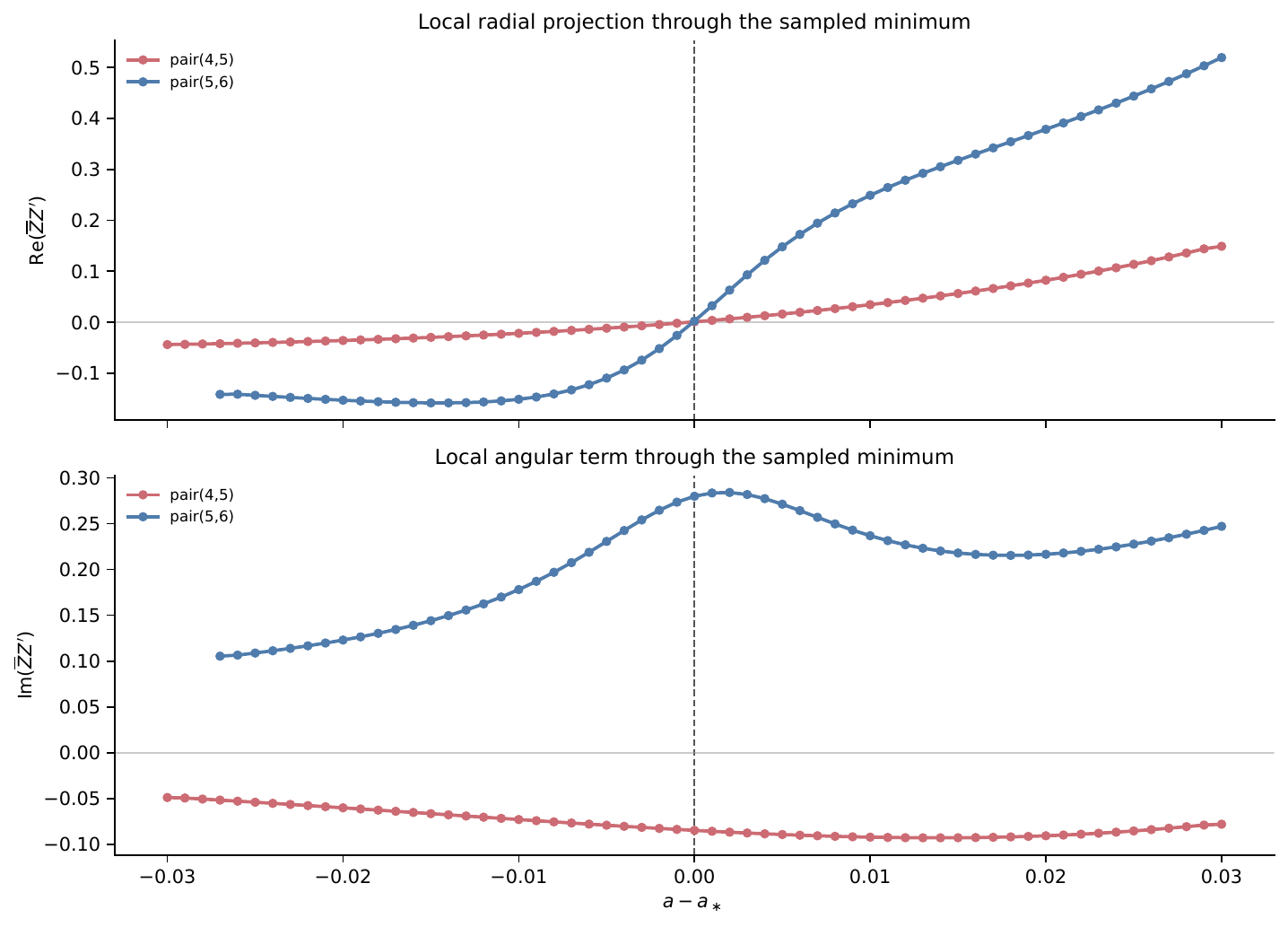}
  \caption{Complex-plane reformulation near the sampled minima for the mainline Kerr case $(s,\ell,m)=(-2,2,2)$. The upper panel shows the local radial projection $\Re(\overline{Z}Z')=rr'$, and the lower panel shows the local angular term $\Im(\overline{Z}Z')=r^2\theta'$, for $\mathrm{pair}(4,5)$ and $\mathrm{pair}(5,6)$. In both cases the radial projection is near zero at the sampled minimum, while the angular term remains nonzero.}
  \label{fig:paper_geometric_reformulation_comparison}
\end{figure}

\smallskip
Near the mainline minima, \(\partial_a g^2\) and \(F\) are examined in Cartesian form and in polar form; figures~\ref{fig:mainline_local_cartesian_balance} and~\ref{fig:paper_geometric_reformulation_comparison} summarize the outcome. The local argument proceeds in two parallel tracks.

\medskip
\noindent\textbf{A. Cartesian balance near the minimum.}
\par\smallskip

Writing \(Z=\Delta\omega_R+\mathrm{i}\Delta\omega_I\) (cf.\ equation~\ref{eq:gap_Z}), equation~(\ref{eq:dg2_master}) becomes
\begin{equation}
  g^2(a)=\Delta\omega_R(a)^2+\Delta\omega_I(a)^2,
\end{equation}
and therefore
\begin{equation}
  \partial_a g^2(a)
  =
  2\Delta\omega_R(a)\,\partial_a\Delta\omega_R(a)
  +
  2\Delta\omega_I(a)\,\partial_a\Delta\omega_I(a).
  \label{eq:partialg}
\end{equation}
We then define the two Cartesian contributions
\begin{align}
  C_R(a) &\equiv 2\Delta\omega_R(a)\,\partial_a\Delta\omega_R(a),
  \label{eq:CR_def}\\
  C_I(a) &\equiv 2\Delta\omega_I(a)\,\partial_a\Delta\omega_I(a),
  \label{eq:CI_def}
\end{align}
so that
\begin{equation}\label{eq:cart_balance}
  \partial_a g^2(a)=C_R(a)+C_I(a).
\end{equation}

Thus a minimum of the scalar gap is not an opaque dip: it is where \(C_R\) and \(C_I\) compete and \(\partial_a g^2\) changes sign, i.e.\ where the real- and imaginary-part channels balance near \(a^*\).

Figure~\ref{fig:mainline_local_cartesian_balance} shows this balance near the minima in the mainline \((s,\ell,m)=(-2,2,2)\) case. Both neighboring pairs share the same pattern. On the left, \(\partial_a g^2<0\) and the gap decreases. Near the minimum, \(C_R\) and \(C_I\) have opposite signs and nearly cancel, so that
\begin{equation}
  \partial_a g^2(a)\approx 0.
  \label{eq:partialgapprox0}
\end{equation}
On the right, the sum becomes positive and the gap turns upward. The minimum is therefore a local derivative-balance event.

Pair dependence is in \emph{how} that cancellation is achieved: both pairs show the same sign switch, but the relative weights of \(C_R\) and \(C_I\) and the sharpness of the transition differ---one mechanism, different local environments.
The Cartesian split is mainly diagnostic; it highlights which channel drives the switch and leads to the complex-plane form below.

\medskip
\noindent\textbf{B. Complex-plane reformulation.}
\par\smallskip

The same content is summarized compactly: from equation~(\ref{eq:dg2_master}),

\begin{equation}
  \partial_a g^2(a)=2\,\Re\!\left(\overline{Z(a)}\,Z'(a)\right)=2F(a).
\end{equation}

To expose the geometry, we write the separation vector in polar form,

\begin{equation}
    Z(a)=r(a)\,\ee^{\ii\theta(a)}.
\end{equation}

Then

\begin{equation}
    \overline{Z}Z' = r r' + \ii r^2\theta',
\end{equation}

so that

\begin{equation}
    \Re(\overline{Z}Z')=r r',
  \qquad
  \Im(\overline{Z}Z')=r^2\theta'.
\end{equation}

The real part is the instantaneous radial projection. The imaginary part is the instantaneous angular rotation. The minimum is thus a radial turning event: radial compression stalls at the minimum, while rotation in the complex plane can continue.

Figure~\ref{fig:paper_geometric_reformulation_comparison} summarizes the complex-plane reformulation near the sampled minima for the mainline Kerr case $(s,\ell,m)=(-2,2,2)$, comparing the local radial projection and the angular background for $\mathrm{pair}(4,5)$ and $\mathrm{pair}(5,6)$.

Thus the minimum is not merely a cancellation in \(\partial_a g^2\); it is a radial-projection event for the separation vector in the complex-frequency plane.

\begin{enumerate}[(i)]
    \item By equations~(\ref{eq:partialg})--(\ref{eq:CI_def}) and equation~(\ref{eq:partialgapprox0}), the Cartesian sum $C_R(a)+C_I(a)$ is exactly the same derivative-level object as $2\Re(\overline{Z(a)}\,Z'(a))$, up to numerical roundoff.
    \item For both pairs, the sampled minimum corresponds to a near-vanishing radial projection in the local complex separation plane.
    \item The angular term remains nonzero at the sampled minimum, so the local event is not a complete halt of the separation vector. Rather, radial compression stalls while angular motion persists.
    \item The two pairs differ not only in where the radial projection crosses through zero, but also in the angular background in which this crossing occurs. In particular, $\mathrm{pair}(5,6)$ is associated with a much larger angular term than $\mathrm{pair}(4,5)$.
\end{enumerate}

\medskip
\noindent\textbf{Full-scan consistency, zero-crossing drift, and representative validation.}
\anchorlabel{subsec:results-global}{R3}
\par\smallskip
Near \(a^*\), the balance admits either the Cartesian split or \(F(a)=\Re(\overline{Z(a)}\,Z'(a))\). The remaining step is to carry this local picture along the full spin scan and to relate different pairs' crossing locations.

\medskip
\noindent\textbf{A. Global consistency of the complex-plane formulation.}
\par\smallskip

As a consistency check, \(\partial_a g^2\), its Cartesian form \(C_R+C_I\), and \(2\Re(\overline{Z}Z')\) are compared over \(a\in[0,0.99]\) for \(\mathrm{pair}(4,5)\) and \(\mathrm{pair}(5,6)\)---they should agree throughout the scan, not only at the minimum.

\begin{figure}[t]
  \centering
  \includegraphics[width=\linewidth]{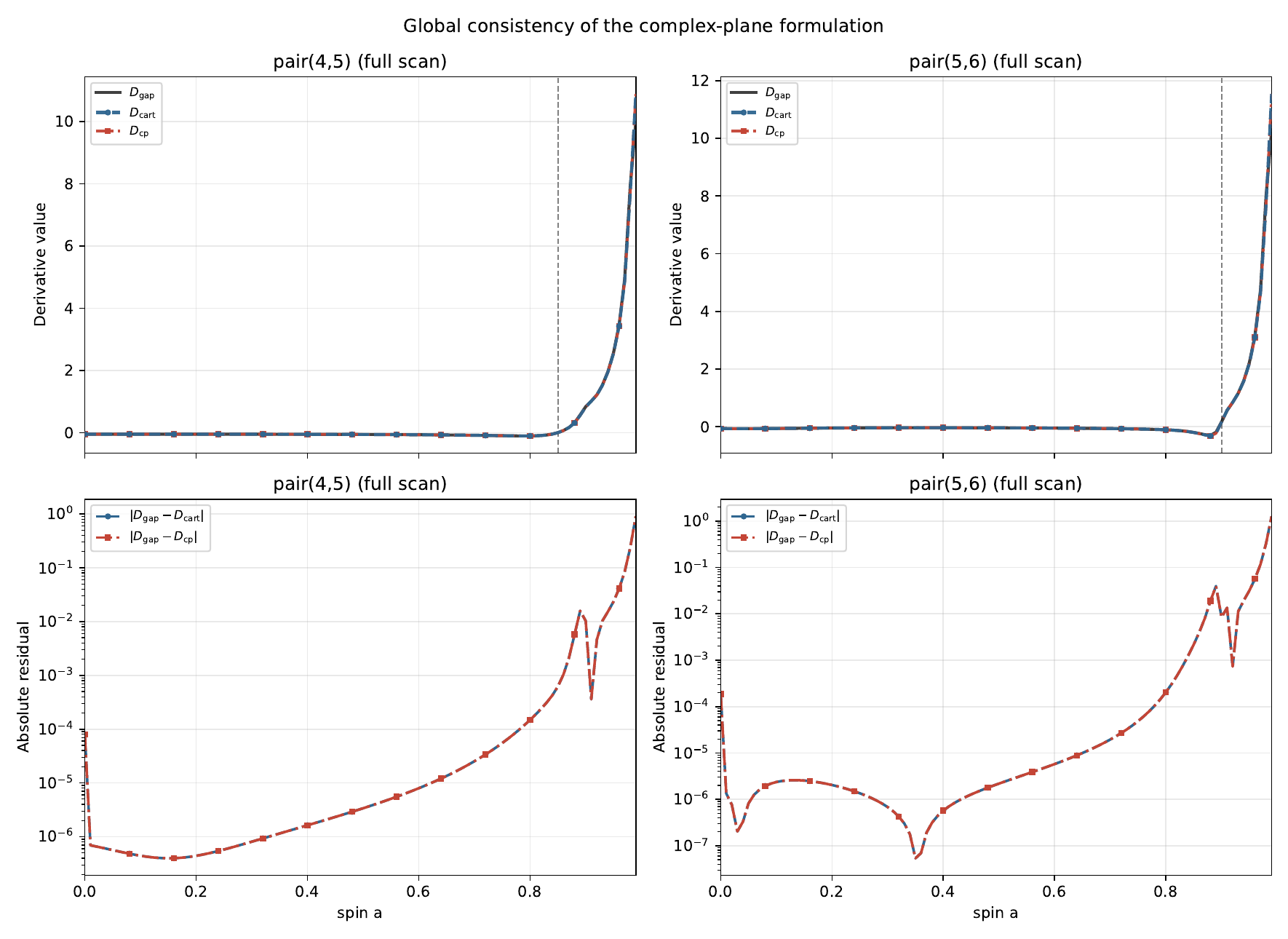}
  \caption{Global consistency of the complex-plane formulation for the mainline Kerr case $(s,\ell,m)=(-2,2,2)$. 
  Top: full-scan comparison of the derivative-level object $D_{\rm gap}=\partial_a g^2$, its Cartesian form $D_{\rm cart}=C_R+C_I$, and its complex-plane form $D_{\rm cp}=2\,\Re\!\left(\overline{Z(a)}\,Z'(a)\right)$ for $\mathrm{pair}(4,5)$ and $\mathrm{pair}(5,6)$. 
  Bottom: absolute residuals $|D_{\rm gap}-D_{\rm cart}|$ and $|D_{\rm gap}-D_{\rm cp}|$ on a logarithmic scale. 
  The dashed vertical line marks the sampled minimum location $a_*$. 
  The near overlap of the three full-scan curves shows that the Cartesian minimum-setting balance and the complex-plane reformulation remain globally consistent along the scan.}
  \label{fig:global_consistency_fullscan}
\end{figure}

Figure~\ref{fig:global_consistency_fullscan} shows the three tracks overlapping with small residuals, so the Cartesian and complex-plane formulations are one and the same mechanism.
Physically, the sampled minimum is the dominant radial turning event along the scan, but each pair sits in a different global background---gentle versus sharp crossings of the radial projection and different angular sectors---so pair dependence appears both locally and globally.
With consistency established, part~B below addresses how drift arises when pairs cross that event at different spins.

\begin{table}[t]
  \centering
  \caption{Restricted representative validation of the minimum-setting and dominant-zero-crossing picture. The table includes same-family positives, external positive transfers, and no-trigger controls. Its purpose is not to claim a family-wide law, but to summarize a small representative set of neighboring-pair cases used to test transferability and selectivity. All six positive validation cases satisfy $|a^{\rm zero}-a^*|<0.02$, i.e.\ within one coarse scan step. Additional broader-scan cases, including lower-overtone candidates, are documented separately in the appendix and are not promoted here to full representative validation.}
  \label{tab:restricted_validation}
  \small
  \renewcommand{\arraystretch}{1.18}
  \setlength{\tabcolsep}{4pt}
  \begin{adjustbox}{max width=\linewidth,center}
  \begin{tabular}{ccccccc}
    \toprule
    $(s,\ell,m)$ & Pair & Interior minimum & Sampled $a^*$ & Dominant $a^{\rm zero}$ & $a^{\rm zero}-a^*$ & Validation status \\
    \midrule
    \multicolumn{7}{l}{\textit{Selected same-family continuations in $(-2,2,0)$}}\\
    \addlinespace[2pt]
    $(-2,2,0)$ & $\mathrm{pair}(4,5)$ & yes & 0.840 & 0.837135 & $-0.002865$ & full local-language transfer \\
    $(-2,2,0)$ & $\mathrm{pair}(5,6)$ & yes & 0.800 & 0.801255 & $+0.001255$ & strong same-family transfer \\
    $(-2,2,0)$ & $\mathrm{pair}(6,7)$ & yes & 0.780 & 0.767833 & $-0.012167$ & edge/distorted transfer with caution \\
    \addlinespace[3pt]
    \midrule
    \multicolumn{7}{l}{\textit{External positive transfers}}\\
    \addlinespace[2pt]
    $(-2,2,2)$ & $\mathrm{pair}(5,6)$ & yes & 0.897 & 0.896911 & $-8.9\times10^{-5}$ & original-mainline strong positive \\
    $(-2,3,1)$ & $\mathrm{pair}(4,5)$ & yes & 0.960 & 0.956078 & $-0.003922$ & clean transfer positive \\
    $(-2,5,2)$ & $\mathrm{pair}(6,7)$ & yes & 0.960 & 0.958944 & $-0.001056$ & moderate positive, cautious transfer \\
    \addlinespace[3pt]
    \midrule
    \multicolumn{7}{l}{\textit{Control; no-trigger}}\\
    \addlinespace[2pt]
    $(-2,2,1)$ & $\mathrm{pair}(2,3)$ & no & --- & --- & --- & smooth control, no trigger \\
    $(-2,4,1)$ & $\mathrm{pair}(0,1)$ & no & --- & --- & --- & low-overtone smooth control \\
    \bottomrule
  \end{tabular}
  \end{adjustbox}
\end{table}

\medskip
\noindent \textbf{B. Pair-dependent zero-crossing drift.}
\par\smallskip
The drift statement can now be made precise. An interior minimum satisfies
\[
F_p(a_p^*)\approx 0,
\qquad
F_p(a)\equiv \Re\!\left(\overline{Z_p(a)}\,Z_p'(a)\right),
\]
so pair-dependent minimum drift is the drift of the dominant zero-crossing location of \(F_p(a)\).

Equivalently, one tracks where the radial projection of the separation changes sign rather than where the scalar gap dips; different pairs reach that turning point at different spins.

For both mainline pairs, the sampled minimum sits very near the dominant zero of \(F_p(a)\):
\[
a^{\rm zero}_{4,5}-a^*_{4,5}=-3.27\times10^{-4},
\qquad
a^{\rm zero}_{5,6}-a^*_{5,6}=-8.9\times10^{-5},
\]
well inside the coarse grid step, so the minimum is anchored to that zero crossing and its drift follows the drift of the dominant radial zero.

The two pairs approach this event differently. The local slope of \(F(a)\) at the crossing is much smaller for \(\mathrm{pair}(4,5)\) than for \(\mathrm{pair}(5,6)\),
\[
\partial_a F|_{4,5}\approx 2.71,
\qquad
\partial_a F|_{5,6}\approx 28.36.
\]
Thus \(\mathrm{pair}(5,6)\) crosses through zero sharply, while \(\mathrm{pair}(4,5)\) crosses gently. This difference in local crossing structure is part of what makes the minimum locations visibly pair-dependent.

Within this framework, drift is not an extra effect: it is the pair dependence of where \(F_p(a)\) crosses zero.

\medskip
\noindent \textbf{C. Restricted representative validation.}
\par\smallskip

To probe how far the picture extends beyond the mainline pairs, a small, deliberately chosen set of additional neighboring-pair cases is examined---without claiming a family-wide law, only whether the minimum-setting and dominant-zero-crossing description transfers in a restricted but nontrivial class.

The first group continues within \((s,\ell,m)=(-2,2,0)\): besides \(\mathrm{pair}(4,5)\), \(\mathrm{pair}(5,6)\) and \(\mathrm{pair}(6,7)\) also show interior minima at
\[
a^*_{4,5}=0.84,\qquad a^*_{5,6}=0.80,\qquad a^*_{6,7}=0.78.
\]
All three match a dominant zero-crossing reading; the first two are clean transfers, while \(\mathrm{pair}(6,7)\) is more distorted and is treated as an edge case.

A second group shows transfer across \((s,\ell,m)\) families, with clean positives and one moderate case near high spin (hence more cautious wording). Across all six positive cases, the sampled minimum and the dominant zero stay aligned within one coarse step, \(|a^{\rm zero}-a^*|<0.02\); the most distorted positive, \(({-2},2,0)\), \(\mathrm{pair}(6,7)\), still satisfies \(|a^{\rm zero}-a^*|=0.012167<0.02\).
The agreement is therefore quantitative, not merely qualitative.

Selectivity is preserved: the control \(({-2},2,1)\), \(\mathrm{pair}(2,3)\), lacks a robust interior minimum, which shows the mechanism is not forced on arbitrary smooth gaps.

In short, the minimum-setting and dominant-zero-crossing picture is not limited to the original mainline pairs; it carries to a restricted validated class and rejects smooth no-trigger cases, without supporting a universal law over the full Kerr spectrum.

\noindent \textbf{\em Discussion.}
\anchorlabel{sec:discussion}{IV}
This section interprets the findings of section~\ref{sec:results}: first by situating the diagnostic \(F\) in standard non-Hermitian and complex-spectral language for Kerr flows (subsection~\ref{subsec:nonhermitian-context}), then by outlining ringdown-oriented implications and an empirical trigger taxonomy from the appendix scans (section~\ref{sec:pheno-triggers}), and finally by stating the main conclusions.

\medskip
\noindent\textbf{Spectral proximity and distinction from exceptional points.}
\anchorlabel{subsec:nonhermitian-context}{D1}
\par\smallskip

Away from isolated zeros of \(Z(a)\), the condition \(F(a)=0\) is equivalent to a stationary point of the modulus \(|Z(a)|\) along the spin direction, because \(\partial_a |Z|^2=2F\) [equation~(\ref{eq:dg2_master})].
This should not be conflated with an exceptional point (EP) in the strict parameter-space sense: EPs require coalescing eigenvalues and nontrivial Jordan-block structure of the underlying spectral problem~\cite{DiasEtAl2022,CavalcanteEtAl2024PRD,CavalcanteEtAl2024PRL,Heiss2012}.
Here \(Z\) is the separation between two \emph{distinct}, label-fixed modes; \(F\) tracks how that separation expands or contracts under infinitesimal changes in \(a\).

The same derivative-level object nonetheless aligns naturally with complex-plane descriptions of Kerr QNM flows, including eigenvalue repulsion and avoided restructuring of nearby poles as external parameters vary~\cite{YangEtAl2013,YangEtAl2013b,YangBertiFranchini2025}.
In that language, an interior minimum of \(g(a)=|Z(a)|\) marks a spin where the neighboring pair is locally closest while \(\Im(\bar Z Z')\) may remain nonzero, i.e.\ radial stall without stopping angular drift of the separation vector.
Pair-dependent drift of \(a^*\) is then the drift of where this radial stall occurs along \(a\) for different overtone pairs.
We use this connection only to anchor the raw gap construction in established non-Hermitian-spectral terminology, not to claim that every triggered minimum coincides with a documented EP crossing.

\medskip
\noindent \textbf{\em Phenomenological implications and empirical trigger taxonomy.}
\anchorlabel{sec:pheno-triggers}{D2}

\medskip
\noindent\textbf{A. Ringdown-relevant phenomenology.}
\par\smallskip

Black-hole spectroscopy and waveform modeling increasingly stress resolving several damped modes at early times, when higher overtones can matter near peak strain~\cite{LondonEtAl2014,BhagwatEtAl2018StartTime,GieslerEtAl2019,Bhagwat2020,BertiEtAl2025Spectroscopy}.
If a neighboring pair develops an interior minimum of \(g_{n,n+1}(a)\), three qualitative points follow from the local geometry: (i)~tighter complex-plane separation can worsen mode confusion or bias in short-window linear fits that assume well-separated poles; (ii)~because the minimum rests on a cancellation between \(\Delta\omega_R\,\partial_a\Delta\omega_R\) and \(\Delta\omega_I\,\partial_a\Delta\omega_I\) [equations~(\ref{eq:CR_def})--(\ref{eq:cart_balance})], modest spin uncertainty maps unevenly to gap error on either side of \(a^*\); (iii)~since \(\Im(\bar Z Z')\) need not vanish at the sampled minimum, phase-sensitive diagnostics need not agree with \(|Z|\) on ``how close'' the two modes are at a given \(a\).

These remarks are qualitative only (no Fisher or likelihood analysis here), but they motivate treating triggered high-overtone pairs as spin-dependent proximity bands: spin ranges where templates and pipelines should allow nearby poles to be nearly degenerate until the data separate them.

\medskip
\noindent\textbf{B. Empirical trigger taxonomy from the appendix scans.}
\par\smallskip

The pair-grouped appendix figures suggest a coarse classification of the \emph{sampled} sectors without claiming completeness over the full Kerr spectrum.
\textit{(i) Lower neighboring pairs} \(\mathrm{pair}(0,1)\) through \(\mathrm{pair}(3,4)\) predominantly yield smooth or weakly structured \(g(a)\), providing no-trigger or weak-trigger backgrounds (figure~\ref{fig:app_pair_grouped_low}).
\textit{(ii) Higher neighboring pairs} \(\mathrm{pair}(4,5)\) through \(\mathrm{pair}(6,7)\) concentrate the clearest interior minima and the strongest cross-\((s,\ell,m)\) differentiation (figure~\ref{fig:app_pair_grouped_high}), consistent with ringdown studies that highlight intermediate-to-high overtones at intermediate-to-high spins~\cite{GieslerEtAl2019,Oshita2021}.
\textit{(iii) Validated triggered cases} in table~\ref{tab:restricted_validation} align the sampled \(a^*\) with the dominant zero of \(F\) within one coarse grid step, \(|a^{\rm zero}-a^*|<0.02\).
\textit{(iv) Edge or distorted triggers} arise near high spin or for borderline pairs (e.g.\ \(\mathrm{pair}(6,7)\) in \((-2,2,0)\)), where shallow crossings or competing zeros warrant cautious wording.

The taxonomy organizes selectivity that is already visible in the appendix; it does not replace an exhaustive survey over all \((s,\ell,m)\) and \(n\).

\medskip
\noindent \textbf{\em Conclusion.}
For Kerr neighboring-overtone separations with labels held fixed along a spin scan, the pair-dependent drift of sampled gap minima admits a compact reformulation: with \(Z(a)\) the complex separation between the two modes and
\[
F(a)=\Re\!\bigl(\overline{Z(a)}\,Z'(a)\bigr),
\]
the dominant zero crossing of \(F\) tracks the minimum; geometrically, the event is a radial turning point of the separation vector in the complex-frequency plane, not merely a feature of the scalar gap.

Restricted validation indicates that this local mechanism transfers to a small but nontrivial class of triggered cases yet fails on smooth no-trigger controls, complementing work on resonant excitation, QNM resonances, excitation factors, and exceptional-point-related structure~\cite{Motohashi2025,YangBertiFranchini2025,LoSabaniCardoso2025,CavalcanteEtAl2024PRL}.
Rather than a new global spectral class, the paper isolates a local organizer for how neighboring-pair minima drift in \(a\), relates it to complex-plane proximity language while distinguishing generic minima from strict exceptional-point coalescence (section~\ref{subsec:nonhermitian-context}), and collects ringdown-oriented remarks plus an appendix-based trigger taxonomy (section~\ref{sec:pheno-triggers}).
The scope stays deliberately limited: the drift is a structured local complex-spectral effect rather than a numerical artifact, without a family-wide law over the full Kerr spectrum.

\noindent \textbf{\em Acknowledgements.}
This work is funded by the National Astronomical Observatories of the Chinese Academy of Sciences, Project No.~E4TG6601.
This work has been supported in part by the National Key Research and Development Program of China under Grant No.~2021YFC2203000.

\noindent \textbf{\em Data availability.}
The data are available from the authors upon reasonable request.

\noindent \textbf{\em Ethics statement.}
This study is theoretical and computational. It does not involve human participants, human data or tissue, or animals; accordingly, no ethics committee approval was required.

\noindent \textbf{\em Competing Interests.}
The authors have no competing financial or non-financial interests to disclose.

\appendix

\begin{center}
{\bfseries Appendix: Broader label-fixed scan and selectivity of interior minima\par}
\end{center}

\vspace{0.5em}

The main text highlights a restricted set of neighboring-pair cases for the minimum-setting and dominant-zero-crossing mechanism; this appendix embeds them in broader pair-grouped scans over several \((s,\ell,m)\) sectors.

For compact layout, figures group by neighboring-pair label (not by \((s,\ell,m)\) family): each panel fixes one pair, and curves compare raw gaps across
\begin{align*}
    (-2,2,0), \quad (-2,2,1), \quad (-2,2,2) \\
    (-2,3,1), \quad (-2,4,1), \quad (-2,5,2)
\end{align*}
Dashed vertical lines mark interior minima when present. The figures document the numerical background behind the main-text choices, not a full mechanism scan for every curve.

They also illustrate selectivity: interior minima are not universal---some pairs show robust triggers compatible with the local picture, while others stay smooth, weak, edge-dominated, or untriggered---without elevating every curve to a validation case or claiming a family-wide law.

\subsection{Overview of the broader scan}

Beyond the main-text representative set, additional cases are summarized only coarsely (clear interior minimum, weak or edge-distorted minimum, or smooth no-trigger behavior) to map the wider landscape around the triggered examples.

For readability, the pair-grouped appendix is divided into two summary figures. Figure~A1 collects the lower neighboring pairs
\[
\mathrm{pair}(0,1),\quad \mathrm{pair}(1,2),\quad \mathrm{pair}(2,3),\quad \mathrm{pair}(3,4),
\]
while figure~A2 collects the higher neighboring pairs
\[
\mathrm{pair}(4,5),\quad \mathrm{pair}(5,6),\quad \mathrm{pair}(6,7).
\]
This emphasis on higher neighboring pairs is also qualitatively consistent with earlier Kerr-ringdown studies from waveform-fitting and excitation-factor perspectives: overtones can be important already near the peak strain amplitude, and for intermediate-to-high spins the fourth, fifth, and sixth overtones are among the most easily excited ones~\cite{GieslerEtAl2019,Oshita2021}.

The broader scan confirms two points. The main-text triggered cases are not numerical curiosities. But they are also not generic across all neighboring pairs. In the lower-pair group of figure~A1, most curves are smooth and mainly provide a no-trigger or weak-trigger background, although stronger structure becomes more visible by \(\mathrm{pair}(2,3)\) and \(\mathrm{pair}(3,4)\). In contrast, the higher-pair group of figure~A2 contains the clearest triggered cases and the strongest cross-sector differentiation, including the mainline \((s,\ell,m)=(-2,2,2)\) behavior.

This broader pattern supports the restricted language used in the main text. The minimum-setting and dominant-zero-crossing mechanism is a structured local explanation for triggered cases. It is not a universal statement about all neighboring-pair combinations in the Kerr spectrum. The appendix is therefore meant to display selectivity, transfer, and distortion at a glance.

The selected pair-grouped overview figures are shown in figures~A1 and~A2.

\clearpage

\setcounter{figure}{0}
\renewcommand{\thefigure}{A\arabic{figure}}

\begin{figure}[H]
    \centering
    \includegraphics[width=0.84\textwidth]{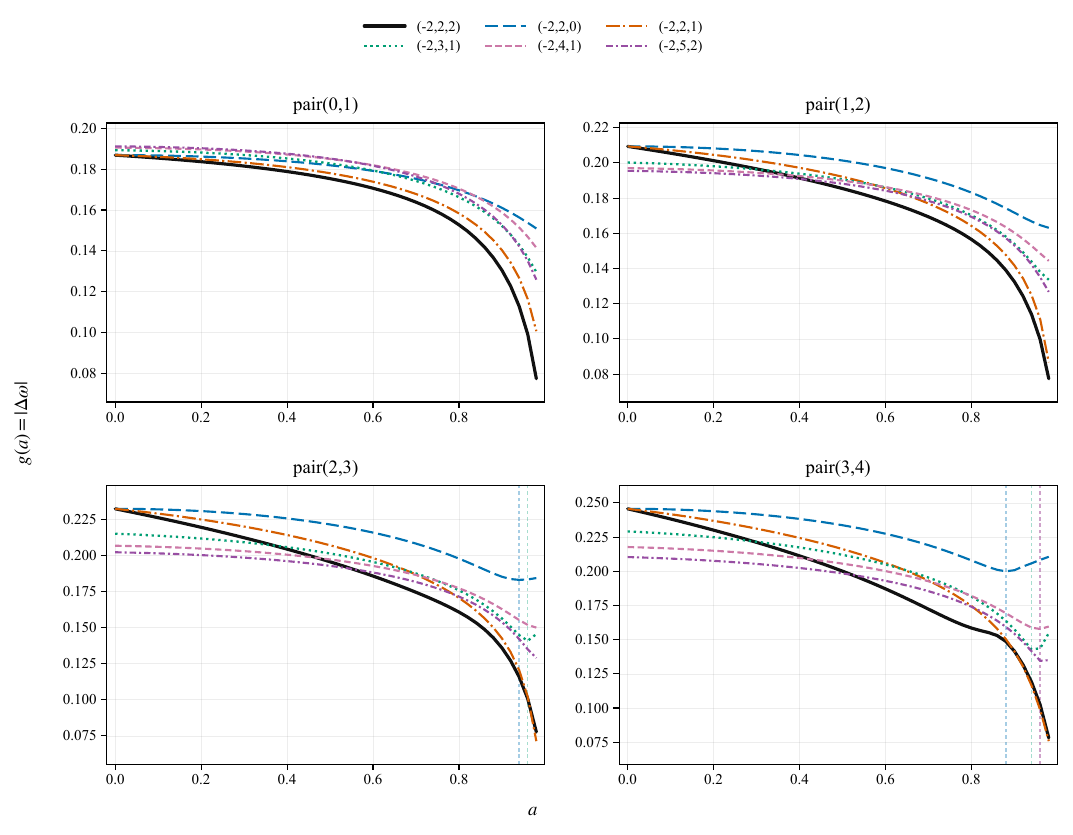}
    \caption{Pair-grouped overview of selected broader label-fixed raw scans for lower neighboring pairs. Each panel fixes one neighboring-pair label, namely \(\mathrm{pair}(0,1)\), \(\mathrm{pair}(1,2)\), \(\mathrm{pair}(2,3)\), or \(\mathrm{pair}(3,4)\), while the overlaid curves compare the corresponding direct raw neighboring-pair gaps across the selected sectors \((s,\ell,m)=(-2,2,0),\,(-2,2,1),\,(-2,2,2),\,(-2,3,1),\,(-2,4,1),\,(-2,5,2)\). Dashed vertical lines mark sampled interior minima when present. This figure mainly provides the low-overtone background, showing that many lower neighboring pairs remain smooth or only weakly structured, even though stronger sector dependence begins to emerge by \(\mathrm{pair}(2,3)\) and \(\mathrm{pair}(3,4)\).}
    \label{fig:app_pair_grouped_low}
\end{figure}

\vspace{-0.8em}

\begin{figure}[H]
    \centering
    \includegraphics[width=0.88\textwidth]{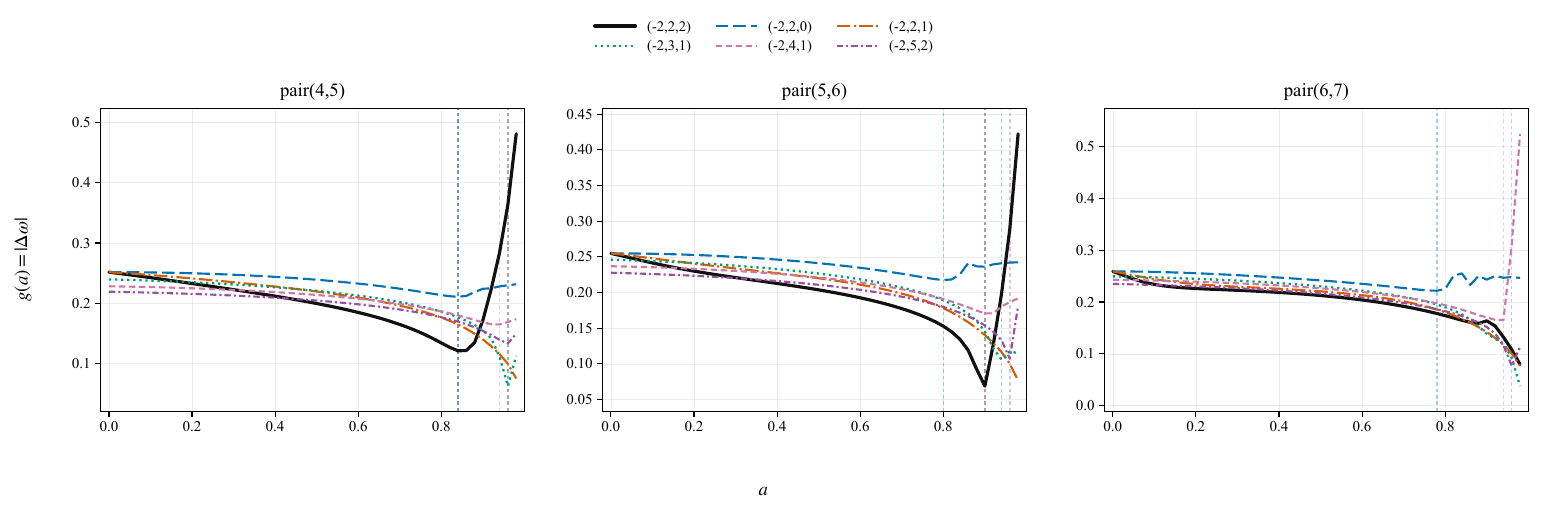}
    \caption{Pair-grouped overview of selected broader label-fixed raw scans for higher neighboring pairs. Each panel fixes one neighboring-pair label, namely \(\mathrm{pair}(4,5)\), \(\mathrm{pair}(5,6)\), or \(\mathrm{pair}(6,7)\), while the overlaid curves compare the corresponding direct raw neighboring-pair gaps across the same selected sectors. Dashed vertical lines mark sampled interior minima when present. This figure contains the clearest triggered cases and the strongest cross-sector differentiation. In particular, the mainline sector \((s,\ell,m)=(-2,2,2)\) exhibits the distinct interior-minimum structure emphasized in the main text, while the other sectors show transfer, distortion, or no-trigger behavior depending on the neighboring pair.}
    \label{fig:app_pair_grouped_high}
\end{figure}

\clearpage

\bibliographystyle{unsrt}
\bibliography{refs}

\begin{thebibliography}{10}

\bibitem{ChandrasekharDetweiler1975}
Subrahmanyan Chandrasekhar and S.~Detweiler.
\newblock {The quasi-normal modes of the Schwarzschild black hole}.
\newblock {\em Proceedings of the Royal Society of London. A. Mathematical and
  Physical Sciences}, 344(1639):441--452, aug 1975.

\bibitem{Chandrasekhar1985}
S.{\~{}}Chandrasekhar.
\newblock {\em {The Mathematical Theory of Black Holes}}.
\newblock Oxford University Press, New York, 1985.

\bibitem{BertiCardosoStarinets2009}
Emanuele Berti, Vitor Cardoso, and Andrei~O Starinets.
\newblock {Quasinormal modes of black holes and black branes}.
\newblock {\em Classical and Quantum Gravity}, 26(16):163001, aug 2009.

\bibitem{BertiEtAl2016KerrSpectroscopy}
Emanuele Berti, Alberto Sesana, Enrico Barausse, Vitor Cardoso, and Krzysztof
  Belczynski.
\newblock {Spectroscopy of Kerr Black Holes with Earth- and Space-Based
  Interferometers}.
\newblock {\em Physical Review Letters}, 117(10):101102, sep 2016.

\bibitem{BertiEtAl2025Spectroscopy}
Emanuele Berti, Vitor Cardoso, Gregorio Carullo, Jahed Abedi, Niayesh Afshordi,
  Simone Albanesi, Vishal Baibhav, Swetha Bhagwat, Jos{\'{e}}~Luis
  Bl{\'{a}}zquez-Salcedo, B{\'{e}}atrice Bonga, Bruno Bucciotti, Giada~Caneva
  Santoro, Pablo~A. Cano, Collin Capano, Mark Ho-Yeuk Cheung, Cecilia Chirenti,
  Gregory~B. Cook, Adrian Ka-Wai Chung, Marina {De Amicis}, Kyriakos Destounis,
  Oscar J.~C. Dias, Walter {Del Pozzo}, Francisco Duque, Will~M. Farr, Eliot
  Finch, Nicola Franchini, Kwinten Fransen, Vasco Gennari, Stephen~R. Green,
  Scott~A. Hughes, Maximiliano Isi, Xisco~Jimenez Forteza, Gaurav Khanna,
  Fech~Scen Khoo, Masashi Kimura, Badri Krishnan, Adrien Kuntz, Macarena Lagos,
  Rico K.~L. Lo, Lionel London, Sizheng Ma, Simon Maenaut, Lorena~Maga{\~{n}}a
  Zertuche, Elisa Maggio, Andrea Maselli, Keefe Mitman, Hayato Motohashi,
  Naritaka Oshita, Costantino Pacilio, Paolo Pani, Rodrigo~Panosso Macedo,
  Chantal Pitte, Lorenzo Pompili, Jaime Redondo-Yuste, Maur{\'{i}}cio Richartz,
  Antonio Riotto, Jorge~E. Santos, Bangalore Sathyaprakash, Laura Sberna,
  Hector~O. Silva, Leo~C. Stein, Alexandre Toubiana, Sebastian~H. V{\"{o}}lkel,
  Julian Westerweck, Huan Yang, Sophia Yi, Nicolas Yunes, and Hengrui Zhu.
\newblock {Black hole spectroscopy: from theory to experiment}, aug 2025.

\bibitem{YangEtAl2013}
Huan Yang, Fan Zhang, Aaron Zimmerman, David~A. Nichols, Emanuele Berti, and
  Yanbei Chen.
\newblock {Branching of quasinormal modes for nearly extremal Kerr black
  holes}.
\newblock {\em Physical Review D}, 87(4):041502, feb 2013.

\bibitem{YangEtAl2013b}
Huan Yang, Aaron Zimmerman, Anıl Zenginoğlu, Fan Zhang, Emanuele Berti, and
  Yanbei Chen.
\newblock {Quasinormal modes of nearly extremal Kerr spacetimes: Spectrum
  bifurcation and power-law ringdown}.
\newblock {\em Physical Review D}, 88(4):044047, aug 2013.

\bibitem{Cardoso2004Rapid}
Vitor Cardoso.
\newblock {Note on the resonant frequencies of rapidly rotating black holes}.
\newblock {\em Physical Review D}, 70(12):127502, dec 2004.

\bibitem{Hod2011NearExtremal}
Shahar Hod.
\newblock {Quasinormal resonances of a massive scalar field in a near-extremal
  Kerr black hole spacetime}.
\newblock {\em Physical Review D}, 84(4):044046, aug 2011.

\bibitem{Motohashi2025}
Hayato Motohashi.
\newblock {Resonant Excitation of Quasinormal Modes of Black Holes}.
\newblock {\em Physical Review Letters}, 134(14):141401, apr 2025.

\bibitem{YangBertiFranchini2025}
Yiqiu Yang, Emanuele Berti, and Nicola Franchini.
\newblock {Black Hole Quasinormal Mode Resonances}.
\newblock {\em Physical Review Letters}, 135(20):201401, nov 2025.

\bibitem{LoSabaniCardoso2025}
Rico K.~L. Lo, Leart Sabani, and Vitor Cardoso.
\newblock {Quasinormal modes and excitation factors of Kerr black holes}.
\newblock {\em Physical Review D}, 111(12):124002, jun 2025.

\bibitem{DiasEtAl2022}
{\'{O}}scar J.~C. Dias, Mahdi Godazgar, Jorge~E. Santos, Gregorio Carullo,
  Walter {Del Pozzo}, and Danny Laghi.
\newblock {Eigenvalue repulsions in the quasinormal spectra of the Kerr-Newman
  black hole}.
\newblock {\em Physical Review D}, 105(8):084044, apr 2022.

\bibitem{CavalcanteEtAl2024PRL}
Jo{\~{a}}o~Paulo Cavalcante, Maur{\'{i}}cio Richartz, and Bruno~Carneiro
  da~Cunha.
\newblock {Exceptional Point and Hysteresis in Perturbations of Kerr Black
  Holes}.
\newblock {\em Physical Review Letters}, 133(26):261401, dec 2024.

\bibitem{CavalcanteEtAl2024PRD}
Jo{\~{a}}o~Paulo Cavalcante, Maur{\'{i}}cio Richartz, and Bruno~Carneiro
  da~Cunha.
\newblock {Massive scalar perturbations in Kerr black holes: Near extremal
  analysis}.
\newblock {\em Physical Review D}, 110(12):124064, dec 2024.

\bibitem{SantosEtAl2026}
Jo{\~{a}}o~S. Santos, Vitor Cardoso, Alexandru Lupsasca, Jos{\'{e}}
  Nat{\'{a}}rio, and Maarten van~de Meent.
\newblock {Resonances in binary extreme-mass-ratio inspirals}.
\newblock {\em Physical Review D}, 113(6):064025, mar 2026.

\bibitem{Heiss2012}
W~D Heiss.
\newblock {The physics of exceptional points}.
\newblock {\em Journal of Physics A: Mathematical and Theoretical},
  45(44):444016, nov 2012.

\bibitem{LondonEtAl2014}
Lionel London, Deirdre Shoemaker, and James Healy.
\newblock {Modeling ringdown: Beyond the fundamental quasinormal modes}.
\newblock {\em Physical Review D}, 90(12):124032, dec 2014.

\bibitem{BhagwatEtAl2018StartTime}
Swetha Bhagwat, Maria Okounkova, Stefan~W. Ballmer, Duncan~A. Brown, Matthew
  Giesler, Mark~A. Scheel, and Saul~A. Teukolsky.
\newblock {On choosing the start time of binary black hole ringdowns}.
\newblock {\em Physical Review D}, 97(10):104065, may 2018.

\bibitem{GieslerEtAl2019}
Matthew Giesler, Maximiliano Isi, Mark~A. Scheel, and Saul~A. Teukolsky.
\newblock {Black Hole Ringdown: The Importance of Overtones}.
\newblock {\em Physical Review X}, 9(4):041060, dec 2019.

\bibitem{Bhagwat2020}
Swetha Bhagwat, Xisco~Jim{\'{e}}nez Forteza, Paolo Pani, and Valeria Ferrari.
\newblock {Ringdown overtones, black hole spectroscopy, and no-hair theorem
  tests}.
\newblock {\em Physical Review D}, 101(4):044033, feb 2020.

\bibitem{BertiCardoso2006}
Emanuele Berti and Vitor Cardoso.
\newblock {Quasinormal ringing of Kerr black holes: The excitation factors}.
\newblock {\em Physical Review D}, 74(10):104020, nov 2006.

\bibitem{DorbandEtAl2006}
Ernst~Nils Dorband, Emanuele Berti, Peter Diener, Erik Schnetter, and Manuel
  Tiglio.
\newblock {Numerical study of the quasinormal mode excitation of Kerr black
  holes}.
\newblock {\em Physical Review D}, 74(8):084028, oct 2006.

\bibitem{ZhangBertiCardoso2013}
Zhongyang Zhang, Emanuele Berti, and Vitor Cardoso.
\newblock {Quasinormal ringing of Kerr black holes. II. Excitation by particles
  falling radially with arbitrary energy}.
\newblock {\em Physical Review D}, 88(4):044018, aug 2013.

\bibitem{Oshita2021}
Naritaka Oshita.
\newblock {Ease of excitation of black hole ringing: Quantifying the importance
  of overtones by the excitation factors}.
\newblock {\em Physical Review D}, 104(12):124032, dec 2021.

\bibitem{OshitaCardoso2024}
N.{\~{}}Oshita and V.{\~{}}Cardoso.
\newblock {Reconstruction of ringdown with excitation factors}.

\bibitem{JaramilloPanossoMacedoAlSheikh2021}
Jos{\'{e}}~Luis Jaramillo, Rodrigo~Panosso Macedo, and Lamis~Al Sheikh.
\newblock {Pseudospectrum and Black Hole Quasinormal Mode Instability}.
\newblock {\em Physical Review X}, 11(3):031003, jul 2021.

\bibitem{Stein2019}
Leo Stein.
\newblock {qnm: A Python package for calculating Kerr quasinormal modes,
  separation constants, and spherical-spheroidal mixing coefficients}.
\newblock {\em Journal of Open Source Software}, 4(42):1683, oct 2019.

\bibitem{Leaver1985}
E.~W. Leaver.
\newblock {An analytic representation for the quasi-normal modes of Kerr black
  holes}.
\newblock {\em Proceedings of the Royal Society of London. A. Mathematical and
  Physical Sciences}, 402(1823):285--298, dec 1985.

\bibitem{CookZalutskiy2014}
Gregory~B. Cook and Maxim Zalutskiy.
\newblock {Gravitational perturbations of the Kerr geometry: High-accuracy
  study}.
\newblock {\em Physical Review D}, 90(12):124021, dec 2014.

\end{thebibliography}

\end{document}